\documentclass[aps,prc,twocolumn,superscriptaddress,nofootinbib]{revtex4-2}

\usepackage{graphicx}   

\begin{document}

\title{Systematic study of hadrons and their quark-component nuclear 
        modification factors}

\author{An-Ke Lei}
\affiliation{Key Laboratory of Quark and Lepton Physics (MOE) and Institute of 
            Particle Physics, Central China Normal University, Wuhan 430079, 
            China}

\author{Dai-Mei Zhou}
\email{zhoudm@mail.ccnu.edu.cn}
\affiliation{Key Laboratory of Quark and Lepton Physics (MOE) and Institute of 
            Particle Physics, Central China Normal University, Wuhan 430079, 
            China}

\author{Yu-Liang Yan}
\email{yanyl@ciae.ac.cn}
\affiliation{China Institute of Atomic Energy, P. O. Box 275 (10), Beijing 
            102413, China}

\author{Du-Juan Wang}
\affiliation{Department of Physics, Wuhan University of Technology, Wuhan 
            430070, China}

\author{Xiao-Mei Li}
\affiliation{China Institute of Atomic Energy, P. O. Box 275 (10), Beijing 
            102413, China}

\author{Gang Chen}
\affiliation{Physics Department, China University of Geoscience, Wuhan 430074, 
            China}

\author{Xu Cai}
\affiliation{Key Laboratory of Quark and Lepton Physics (MOE) and Institute of 
            Particle Physics, Central China Normal University, Wuhan 430079, 
            China}

\author{Ben-Hao Sa}
\email{sabh@ciae.ac.cn}
\affiliation{Key Laboratory of Quark and Lepton Physics (MOE) and Institute of 
            Particle Physics, Central China Normal University, Wuhan 430079, 
            China}
\affiliation{China Institute of Atomic Energy, P. O. Box 275 (10), Beijing 
            102413, China}

\date{\today}

\begin{abstract}
    We have systematically studied the connection (correspondence) between 
    hadron and its quark component nuclear modification factors and the 
    flavor (mass) ordering at both parton and hadron levels in the 
    nucleus-nucleus collisions at the LHC energies by the PACIAE model. It 
    turns out that the correspondence and the mass ordering are generally held, 
    irrespective of the rapidity, centrality, reaction energy, and the 
    collision system size. The nuclear modification factors of hadrons in 
    the final hadronic state show clear mass ordering, which 
    should be studied further both theoretically and experimentally.
\end{abstract}

\maketitle


\section{Introduction} \label{sec:intro}
    The hot and dense Quark-Gluon Plasma (QGP), a phase of deconfined nuclear 
    matter, has been found to be created in ultra-relativistic heavy-ion 
    collisions at both the Relativistic Heavy Ion Collider 
    (RHIC)~\cite{PHENIX:2004vcz,BRAHMS:2004adc,PHOBOS:2004zne,STAR:2005gfr} 
    and the Large Hadron Collider 
    (LHC)~\cite{ALICE:2010suc,ATLAS:2010isq,ALICE:2010yje,CMS:2011iwn,ALICE:2012jsl}. 
    One of the most important signatures for QGP formation is the suppression 
    of hadron production at high transverse momentum ($p_T$) due to the energy 
    loss effect (jet quenching)~\cite{Bjorken:1982tu,Gyulassy:1990ye}. To 
    quantify such a suppression effect, the nuclear modification factor 
    measurement was proposed~\cite{Wang:1998bha}.
    It is defined as the ratio of the $p_T$-differential multiplicity 
    $dN/dp_T$ in nucleus-nucleus collisions ($AA$) to the one in 
    nucleon-nucleon collisions ($pp$), scaled by the binary collision number 
    $\left\langle N_{coll} \right\rangle$ for a given range of 
    centrality interval~\cite{Wang:1998bha,Klein-Bosing:2018izr}
    \begin{equation}
        R_{AA}^{X}(p_T)=\frac{1}{ \left\langle N_{coll} \right\rangle }
        \frac{dN_X^{AA}/dp_T}{dN_X^{pp}/dp_T},
        \label{eq:raa}
    \end{equation}
    where $X$ stands for a specific particle, 
    $\left\langle N_{coll} \right\rangle$ can be obtained 
    from the optical Glauber model and/or the Monte-Carlo Glauber 
    model~\cite{Glauber:1970jm,Miller:2007ri,STAR:2008med,ALICE:2013hur,Loizides:2014vua,Loizides:2017ack}.
    The value of $R_{AA}$ would be unity if an $AA$ collision is just
    a simple superposition of the $pp$ collisions. 
    Conversely, one could expect a non-unity $R_{AA}$ in the presence of the 
    cold and hot nuclear medium effects. Therefore, $R_{AA}$ could serve as 
    an excellent observable for exploring the jet quenching effect.

    The QGP medium induced jet quenching effect is expected to suppress the 
    $R_{AA}$ of partons in the final partonic state (FPS). On the other hand, 
    the $R_{AA}$ of hadrons in the final hadronic state (FHS), the one actually 
    measurable in experiments, receives convoluted contributions from the 
    partonic jet quenching effect and the hadronic energy loss effect in the 
    hadronization and hadronic rescattering stage. Naturally, one would seek 
    the connection (correspondence) between the $R_{AA}$ of partons in FPS and 
    the $R_{AA}$ of hadrons in FHS. Taking $R_{AA}$ of $\Lambda$ as an 
    example, the connection between $R_{AA}$ of $\Lambda$ and that of its 
    quark component could be considered in simple forms like follow:
    \begin{enumerate}
        \item connect $R_{AA}^{\Lambda}$ to single $R_{AA}^u$ ($R_{AA}^{d}$,
        $R_{AA}^{s}$);
        \item connect $R_{AA}^{\Lambda}$ to $R_{AA}^{u+d}$ ($R_{AA}^{u+s}$,
        $R_{AA}^{d+s}$), calculated by the sum of $u$- and $d$- 
        ($u$- and $s$-, $d$- and $s$-) quark $p_T$-distributions without 
        weight factor;
        \item connect $R_{AA}^{\Lambda}$ to $R_{AA}^{u+d+s}$, calculated by 
        the sum of $u$-, $d$-, and $s$-quark $p_T$-distributions without 
        weight factor.
    \end{enumerate}
        However, all of them are incomplete:
    \begin{enumerate}
        \item the $d$- and $s$- ($u$- and $s$-, $u$- and $d$-) constituent 
        quarks are ignored;
        \item the $s$- ($d$-, $u$-) constituent quark is excluded;
        \item the contribution of sea quark $s$ is underestimated.
    \end{enumerate}

    As far as we know, a correct connection (correspondence) is unable to be 
    introduced from the first principle theory, even from the recombination 
    (coalescence) model~\cite{Hwa:2002tu,Greco:2003xt,Greco:2003mm,Fries:2003vb,Fries:2003kq,Shao:2004cn}, 
    because of the complication in dealing with the flavor composition of 
    constituent quarks. Recently, a connection (correspondence) between 
    the hadron $R_{AA}$ (in FHS) and its quark component one (in FPS) has been 
    proposed by us for the first time~\cite{Sa:2022pnd}, and the $R_{AA}$ mass 
    ordering at hadron level, initiating from the dead-cone 
    effect~\cite{Dokshitzer:1991fd}, is also explored~\cite{Sa:2022pnd}. In 
    this work, we extend the study to the rapidity, centrality, reaction energy, 
    and the collision system size dependences of above two physical phenomena.

    The paper is organized as follows. In Sec.~\ref{sec:method} the 
    methodology to study the $R_{AA}$ correspondence of the hadron and its 
    quark component is provided, after the introduction section 
    Sec.~\ref{sec:intro}. In Sec.~\ref{sec:res}, the results of 
    correspondence and $R_{AA}$ mass ordering dependent on the rapidity, 
    centrality, reaction energy and the collision system size are presented. 
    We summarize in Sec.~\ref{sec:sum}.

\section{Methodology} \label{sec:method}
    In this work, we follow the formalism of the correspondence between the 
    hadron $R_{AA}^h$ in FHS and its quark component $R_{AA}^{h \textrm{-} q}$ 
    (the script $h \textrm{-} q$ refers to the quark component of the hadron 
    $h$) in FPS established in Ref.~\cite{Sa:2022pnd}.
    The brief physical deduction is described as follows:

    Considering the hadron normalized $p_T$-differential distribution 
    \[
        \frac{1}{N_h}dN_h/dp_T,
    \]
    its corresponding quark component normalized $p_T$-differential 
    distribution is
    \[
        \frac{1}{N_{h \textrm{-} q}} \sum\limits_q \frac{1}{N_q}dN_q/dp_T .
    \]
    In the above expressions, $N_h$ ($N_q$) refers to the multiplicity of the
    hadron $h$ (quark $q$). $N_{h \textrm{-} q}$ denotes the number of 
    constituent quarks in a hadron $h$ and the sum is taken over all 
    constituent quarks.

    Multiplying above two expressions by $N_h$, one can get the hadron 
    un-normalized $p_T$-differential distribution 
    \[
        dN_h/dp_T
    \]
    and the corresponding quark component un-normalized 
    $p_T$-differential distribution
    \[
        \frac{1}{N_{h \textrm{-} q}} \sum\limits_q \frac{N_h}{N_q}dN_q/dp_T .
    \]

    Substituting above two un-normalized $p_T$-differential distributions into 
    equation (\ref{eq:raa}), respectively, one obtains the hadron nuclear 
    modification factor
    \begin{equation}
        R_{AA}^{h}(p_T)=
        \frac{1}{ \left\langle N_{coll} \right\rangle }
        \frac{dN_h^{AA}/dp_T} {dN_h^{pp}/dp_T}, 
        \label{eq:raa_h}
    \end{equation}
    and its corresponding quark component nuclear modification factor
    \begin{equation}
        R_{AA}^{h \textrm{-} q}(p_T) = 
        \frac{1}{ \left\langle N_{coll} \right\rangle } 
        \frac{ \sum \limits_q {w_q^{AA} dN_q^{AA}/dp_T} } 
             { \sum \limits_q { w_q^{pp} dN_q^{pp}/dp_T} }, 
        \label{eq:raa_h_q}
    \end{equation}
    where $w_q = N_h/N_q$ is the weight factor.

    To investigate the correspondence between hadron and its quark component 
    and the mass ordering at hadron level in $R_{AA}$, a numerical Monte-Carlo 
    event generator, PACIAE~\cite{Sa:2011ye}, is employed to simulate the $pp$ 
    and $AA$ collisions. PACIAE is a microscopic parton and hadron cascade 
    model based on the PYTHIA6.4 event generator~\cite{Sjostrand:2006za}.

    For nucleon-nucleon ($NN$) collisions, with respect to PYTHIA, the partonic 
    and hadronic rescatterings are introduced before and after the 
    hadronization, respectively. The final hadronic state is developed from the 
    initial partonic hard scattering and parton showers, followed by parton 
    rescattering, string fragmentation, and hadron rescattering stages. Thus, 
    the PACIAE model provides a multi-stage transport description on the 
    evolution of the $NN$ collision system.

    For $AA$ collisions, the initial positions of nucleons in the colliding nuclei 
    are sampled according to the Woods-Saxon distribution. Together with the 
    initial momentum setup of $p_{x} = p_{y} = 0$ and $p_{z} =p_{\rm beam}$ for 
    each nucleon, a list containing the initial state of all nucleons in a given 
    $AA$ collision is constructed. 

    A collision happened between two nucleons 
    from different nuclei if their relative transverse distance is less than or 
    equal to the minimum approaching distance: 
    $D\leq\sqrt{\sigma_{NN}^{tot}/\pi}$. The collision time is 
    calculated with the assumption of straight-line trajectories. All such 
    nucleon pairs compose an $NN$ collision time list.

    The earliest $NN$ collision 
    in the list will be executed by PYTHIA (PYEVNW subroutine) with the 
    hadronization temporarily turned-off, as well as the strings and diquarks
    broken-up. The nucleon list and $NN$ collision time list are then updated 
    accordingly for the iteration of the next $NN$ collision. By repeating the 
    aforementioned steps till the $NN$ collision list is empty, the initial 
    partonic state is constructed for an $AA$ collision.

    Then, the partonic rescatterings are performed, where the LO-pQCD 
    parton-parton cross section~\cite{Combridge:1977dm,Field1989ApplicationsOP} 
    is employed. After partonic rescattering, the string is recovered and then 
    hadronized with the Lund string fragmentation scheme.

    The Lund string fragmentation is a phenomenological hadronization model. 
    Here the key assumption is the iterative string breaking procedure: 
    Supposing an iterative string breaking process starting from the $q_0$ 
    end of a $q_0 \bar{q_0}$ string, if the string potential energy is large 
    enough, a new $q_1 \bar{q_1}$ pair may be excited from the vacuum, such that 
    a meson $M_1$ of $q_0 \bar{q_1}$ is formed and the $q_1$ quark left behind. 
    Later on, the $q_1$ in its turn may excite a $q_2 \bar{q_2}$ pair from the 
    vacuum and combine into another meson $M_2$ with the $\bar{q_2}$. This 
    breaking process repeats and repeats until the potential energy is not 
    large enough. The similar is for the baryon-antibaryon ($B \bar{B}$) 
    production in the popcorn model. But instead of starting from one end of 
    $q_0 \bar{q_0}$ string, in popcorn model the process is performed on the 
    entire $q_0 \bar{q_0}$ 
    string. Suppose $q_0 \bar{q_0}$ is a red-antired ($r \bar{r}$) string and 
    has enough potential energy to excite three pairs of green-antigreen 
    ($g \bar{g}$), blue-antiblue ($b \bar{b}$), and antiblue-blue ($\bar{b} b$) 
    iteratively from the vacuum in between the $r \bar{r}$ string. Then a 
    red-green-blue and antiblue-antigreen-antired baryon-antibaryon 
    ($B \bar{B}$) pair may form together with a antiblue-blue meson ($M$) in 
    between the original $r \bar{r}$ pair, resulting in a ($B M \bar{B}$) 
    configuration~\cite{Sjostrand:2006za,Bierlich:2022pfr}.

    Taking the meson production as an example, once the $q_{i-1}$ and
    $\bar{q_i}$ flavors are sampled, a selection should be made between the
    possible multiplets. The different multiplets have different relative
    composition probabilities, which are not given by first principle but 
    depend on the fragmentation processes, 
    cf. Refs.~\cite{Sjostrand:2006za,Bierlich:2022pfr} for the details.

    Finally, the above formed intermediate hadronic state proceeds into the 
    hadronic rescattering stage and produces the final hadronic state observed 
    in the experiments.

    Thus PACIAE Monte-Carlo simulation provides a complete description of the $NN$ 
    and/or $AA$ collisions, which includes the partonic initialization stage, 
    partonic rescattering stage, hadronization stage, and the hadronic 
    rescattering stage. Meanwhile, the PACIAE model simulation could be requested 
    to stop at any of the above stages conveniently. In this work, the simulations 
    are stopped at the final partonic state (FPS) after partonic rescattering 
    or at final hadronic state (FHS) after hadronic rescattering for the 
    calculations of $R_{AA}^{h \textrm{-} q}$ and $R_{AA}^h$, respectively. 
    More details could be found in the Ref.~\cite{Sa:2011ye}.

    In order to be self-consistent, the tuning parameters are kept the same as 
    those in Ref.~\cite{Sa:2022pnd}: A factor multiplying on the hard scattering 
    cross-section $K$=2.7 (0.7), the Lund string fragmentation parameters of 
    $\alpha$=1.3 (0.1) and $\beta$=0.09 (0.58), as well as the Gaussian width 
    of the primary hadron transverse momentum distribution $\omega$=0.575 
    (0.36) are implemented in $AA$ ($pp$) simulations.

\section{Results and discussions} \label{sec:res}
    In Ref.~\cite{Sa:2022pnd}, we have proposed a method connecting the 
    hadron nuclear modification factor $R_{AA}^h$ in FHS to its quark component 
    nuclear modification factor $R_{AA}^{h \textrm{-} q}$ in FPS, and explored 
    the mass ordering in nuclear modification factor at both parton and 
    hadron levels in the 0-5\% most central 
    Pb+Pb collisions at $\sqrt{S_{NN}}$=2.76 TeV. In this section, we will 
    expand our research discussing the dependences of $R_{AA}$  correspondence 
    and $R_{AA}$ mass ordering on rapidity, centrality, 
    reaction energy, and the collision system size. 
    Our simulations are all performed in full $\eta$ phase space, except the 
    in Sec.~\ref{subsec:eta}.

\subsection{Rapidity dependence} \label{subsec:eta}
\begin{figure}[htbp]
        \centering
        \includegraphics[width=0.5\textwidth]{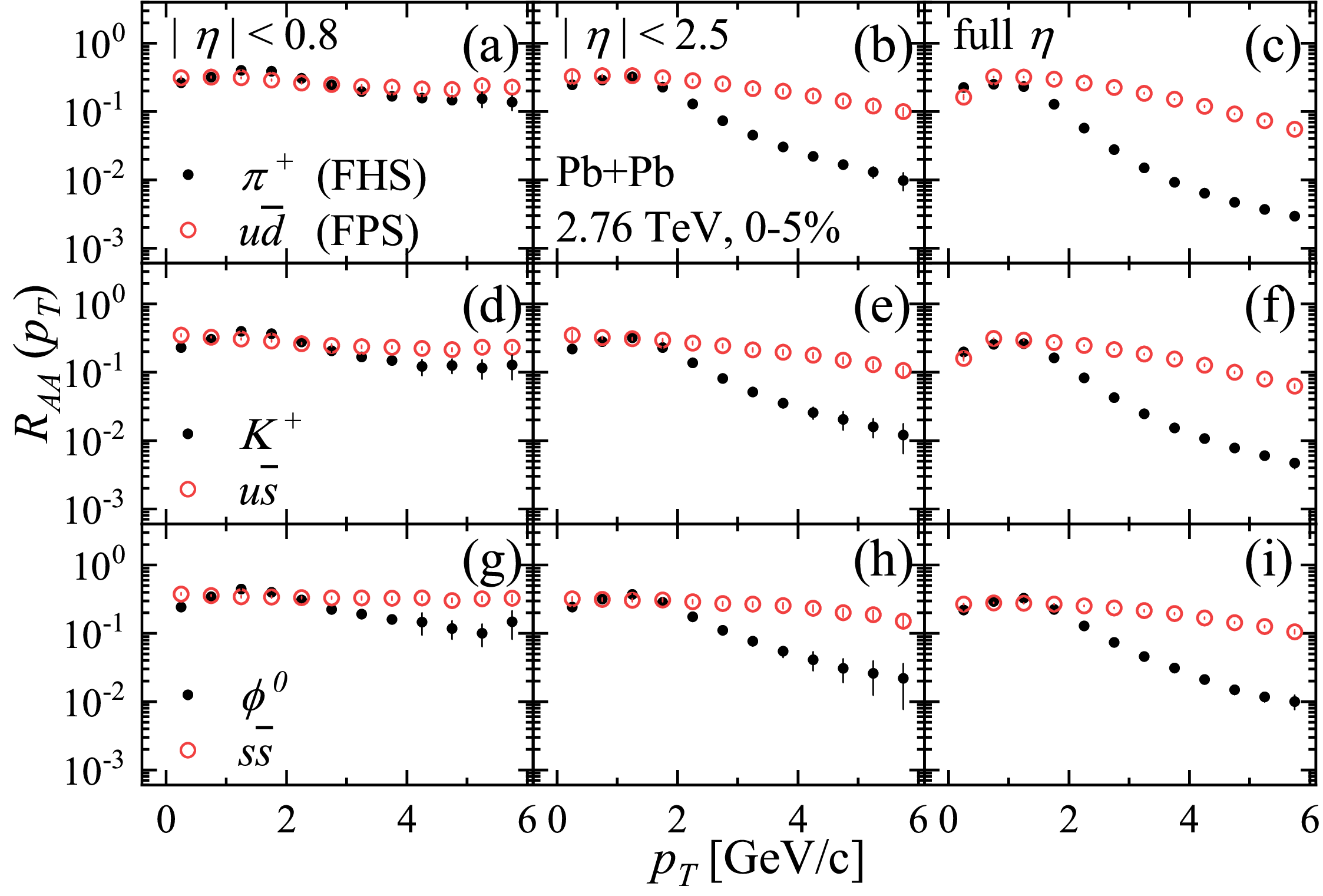}
        \caption{The 
        simulated correspondence between $R_{AA}$ of mesons (in FHS, black 
        solid circles) and their quark component (in FPS, red open circles) 
        in the 0-5\% most central Pb+Pb collisions at $\sqrt{S_{NN}}$=2.76 TeV
        within three pseudo-rapidity ranges.}
        \label{fig:wRaa_mes_diff_eta}
    \end{figure}

\begin{figure}[htbp]
        \centering
        \includegraphics[width=0.5\textwidth]{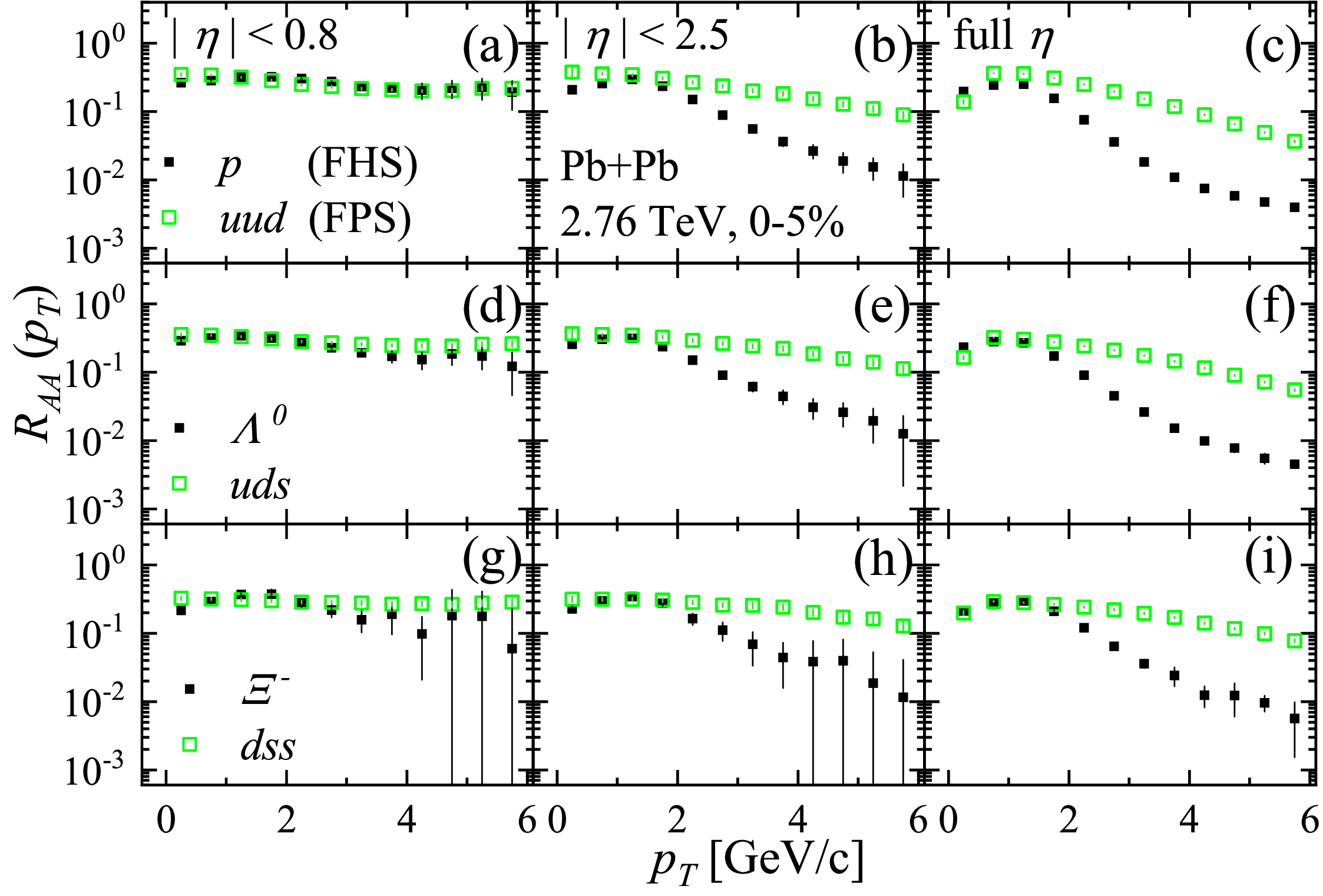}
        \caption{The 
        simulated correspondence between $R_{AA}$ of baryons (in FHS, black 
        solid squares) and their quark component (in FPS, green open squares) 
        in the 0-5\% most central Pb+Pb collisions at $\sqrt{S_{NN}}$=2.76 TeV 
        within three pseudo-rapidity ranges.}
        \label{fig:wRaa_bar_diff_eta}
    \end{figure}

    Fig.~\ref{fig:wRaa_mes_diff_eta} shows the meson $R_{AA}$ in FHS (black 
    solid circles) and its quark component $R_{AA}$ in FPS (red open circles) 
    within different $\eta$ ranges in the 0-5\% most central Pb+Pb collisions 
    at $\sqrt{S_{NN}}$=2.76 TeV. From top to bottom rows are $R_{AA}$ of 
    $\pi^+ (u\overline{d})$, $K^+ (u\overline{s})$ and 
    $\phi^0 (s\overline{s})$, while from left to right columns are those in 
    $|\eta|<0.8$, 2.5 and full $\eta$ phase space, 
    respectively~\footnote{Such a graph 
    arrangement manner would also be utilized hereafter, just for different 
    particle sectors and conditions applied.}. 
    Fig.~\ref{fig:wRaa_bar_diff_eta} shows the same content, just for the 
    baryon sector. A peak appears at $p_T $ around $1 \sim 2$ GeV/c, 
    which is the so-called Cronin effect~\cite{Cronin:1974zm} attributed 
    to the multiple scattering of initial partons~\cite{Wang:1998ww}. 

    In Fig.~\ref{fig:wRaa_mes_diff_eta} and~\ref{fig:wRaa_bar_diff_eta}, 
    one can see that the correspondence between hadron $R_{AA}$ and its quark 
    component $R_{AA}$ exists in all three rapidity ranges. The meson and 
    baryon $R_{AA}$ are smaller than their quark component $R_{AA}$ in the 
    $p_T$ region above $p_T \approx $ 2 GeV/c, even in the midrapidity interval 
    of $|\eta|<0.8$. This is due to the additional energy loss experienced by 
    the final state hadrons in the hadronization and hadronic rescattering 
    stages, while their quark components only experience partonic energy 
    losses. Moreover, this discrepancy becomes more pronounced as the $\eta$ 
    range increases. This is because the energy loss increases with the 
    increasing number of colliding and radiating particles involved in a wider 
    $\eta$ range.

\begin{figure}[htbp]
        \centering
        \includegraphics[width=0.5\textwidth]{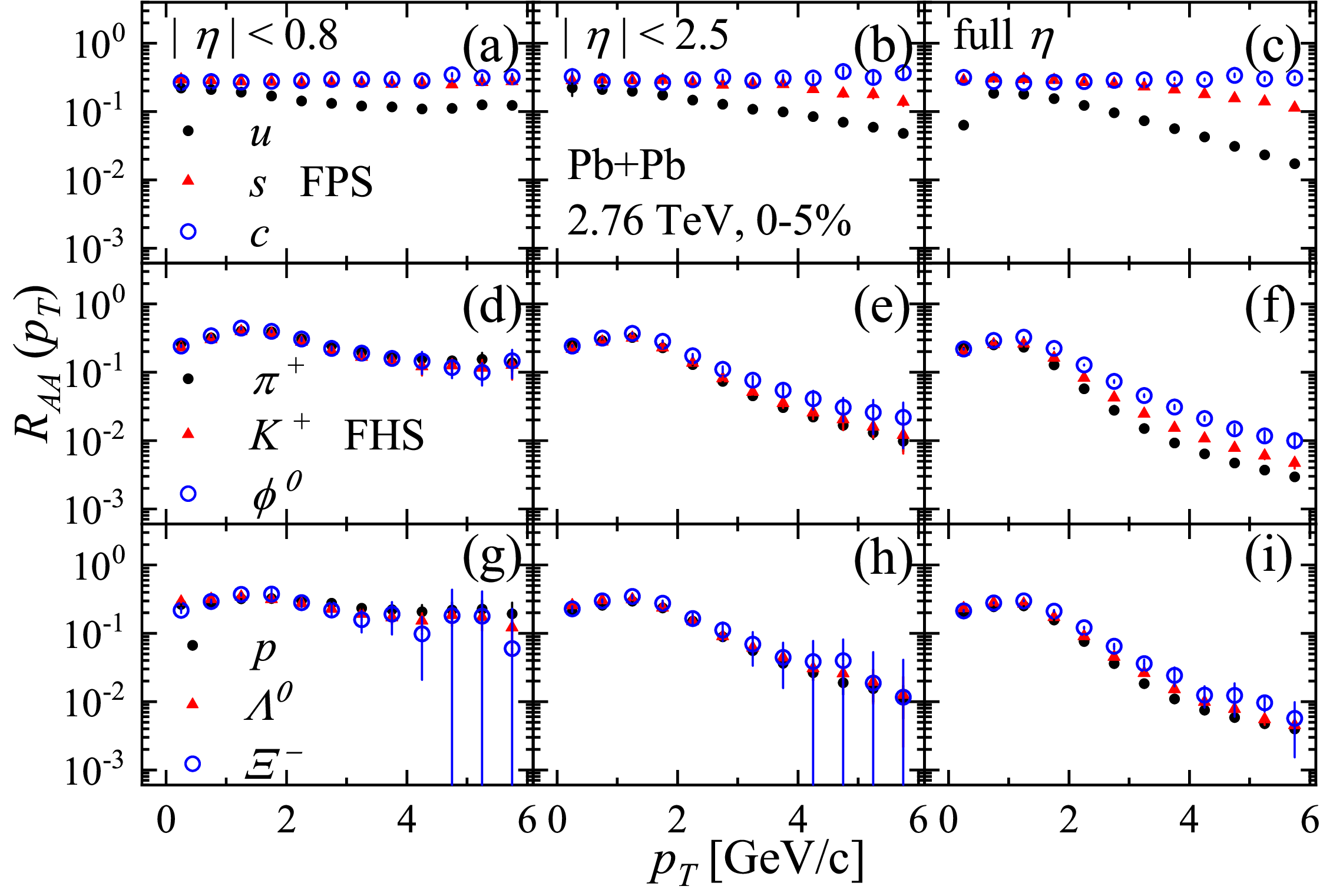}
        \caption{The 
        simulated $R_{AA}$ of quarks (in FPS), mesons 
        (in FHS) and baryons (in FHS) in the 0-5\% most central Pb+Pb collisions 
        at $\sqrt{S_{NN}}$=2.76 TeV within three pseudo-rapidity ranges.}
        \label{fig:raa_diff_eta}
    \end{figure}

    In Fig.~\ref{fig:raa_diff_eta}, we compare the mass ordering of 
    the quarks (in FPS), mesons (in FHS), and the baryons (in FHS) nuclear 
    modification factors among three different $\eta$ phase spaces, as 
    displayed from left to right column, respectively. The mass ordering seems 
    to be generally held for quarks and mesons. However, for baryons, the 
    mass ordering is only recognized in full $\eta$ phase space. This can be 
    explained by the relative mass discrepancy among the selected particle 
    species. The relative mass difference among baryons 
    ($ m_{p} \approx 0.938 \textrm{~GeV}, m_{\Lambda^0} 
    \approx 1.116 \textrm{~GeV~} \textrm{and~} m_{\Xi^-} \approx 1.322 
    \textrm{~GeV} $~\cite{ParticleDataGroup:2020ssz}) is smaller than that 
    among the mesons ($ m_{\pi^+} \approx 140 \textrm{~MeV}, 
    m_{K^+} \approx 494 \textrm{~MeV~} \textrm{and~}m_{\phi^0} \approx 1.02 
    \textrm{~GeV} $) and much smaller than that among the quarks 
    ($ m_u \approx 2.2 \textrm{~MeV}, m_s \approx 93 \textrm{~MeV~} 
    \textrm{and~} m_c \approx 1.27 \textrm{~GeV} $). For quark sector, 
    a nearly flat $R_{AA}$ of the heavy $c$-quark is observed. 
    This could be understood from the fact that the $c$-quark is produced in 
    initial hard processes and transparent in the partonic and hadronic 
    rescatterings. Hence the $p_T$ distribution of $c$-quark in Pb+Pb 
    collisions is approximately parallel to that in p+p collisions at the same 
    energy.

\subsection{Centrality dependence} \label{subsec:cent}
\begin{figure}[htbp]
        \centering
        \includegraphics[width=0.5\textwidth]{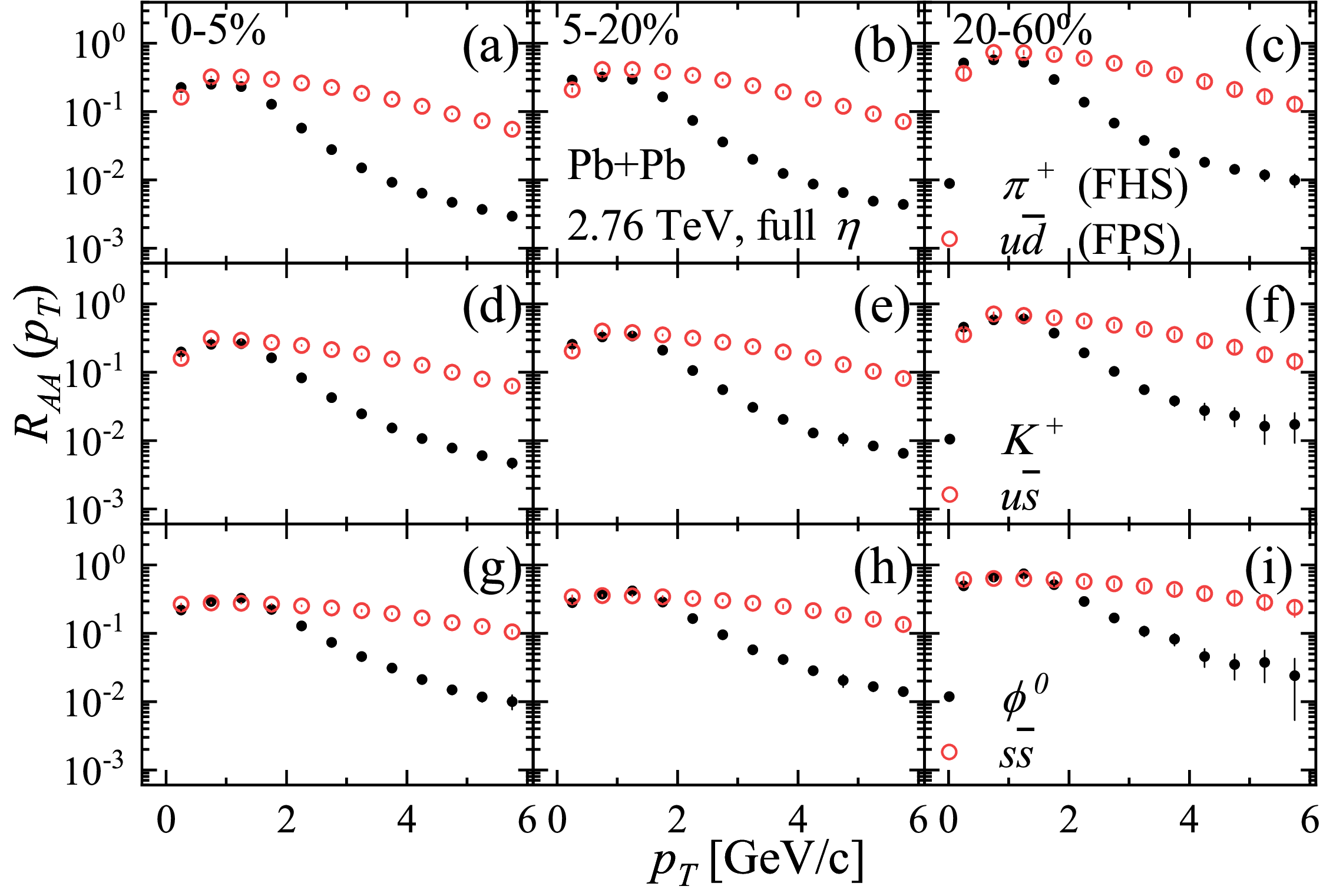}
        \caption{The 
        simulated correspondence between $R_{AA}$ of mesons (in FHS, black 
        solid circles) and their quark component (in FPS, red open circles) in 
        the different centrality classes of Pb+Pb collisions at 
        $\sqrt{S_{NN}}$=2.76 TeV.}
        \label{fig:wRaa_mes_diff_cent}
    \end{figure}

\begin{figure}[htbp]
        \centering
        \includegraphics[width=0.5\textwidth]{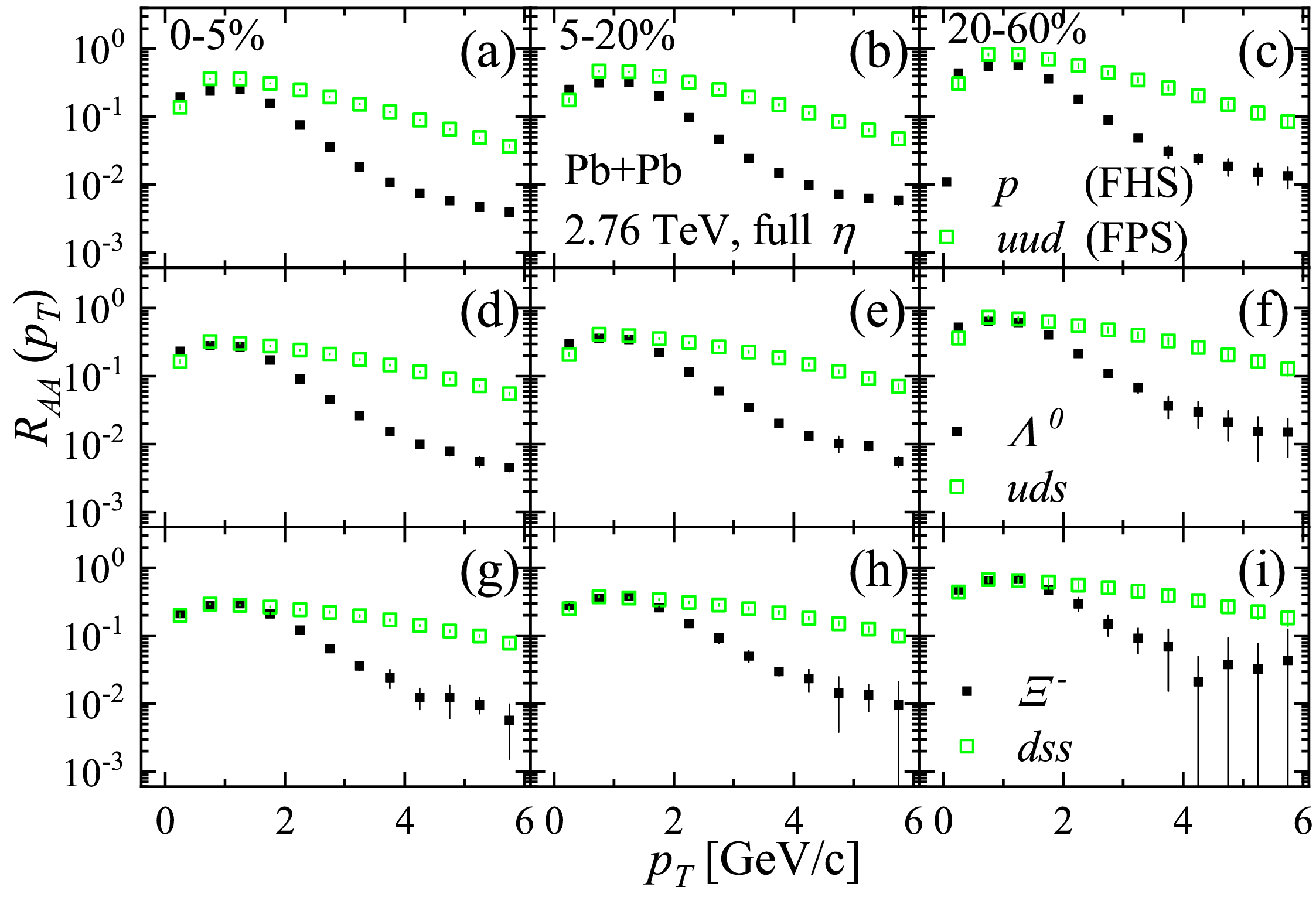}
        \caption{The 
        simulated correspondence between $R_{AA}$ of baryons (in FHS, 
        black solid squares) and 
        their quark component (in FPS, green open squares) in the different 
        centrality classes of Pb+Pb 
        collisions at $\sqrt{S_{NN}}$=2.76 TeV.}
        \label{fig:wRaa_bar_diff_cent}
    \end{figure}

    In Figs~\ref{fig:wRaa_mes_diff_cent} and \ref{fig:wRaa_bar_diff_cent}, 
    we show the simulated correspondence between hadron (meson and baryon in 
    FHS) and its quark component (in FPS) in $R_{AA}$ in the 0-5\%, 5-20\% and 
    20-60\% centrality classes Pb+Pb collisions at $\sqrt{S_{NN}}$=2.76 TeV. 
    Still, the more stronger suppresion of the hadron $R_{AA}$ than its quark 
    component $R_{AA}$ can be found in all three centrality classes. One can 
    see approximately a more depressed magnitude of both $R_{AA}$ in more 
    central centrality class stemming from a stronger hot medium effect. 
    Nevertheless, the discrepancy between hadron $R_{AA}$ and its quark 
    component $R_{AA}$ brought about by centrality classes is not so 
    significant.

\begin{figure}[htbp]
        \centering
        \includegraphics[width=0.5\textwidth]{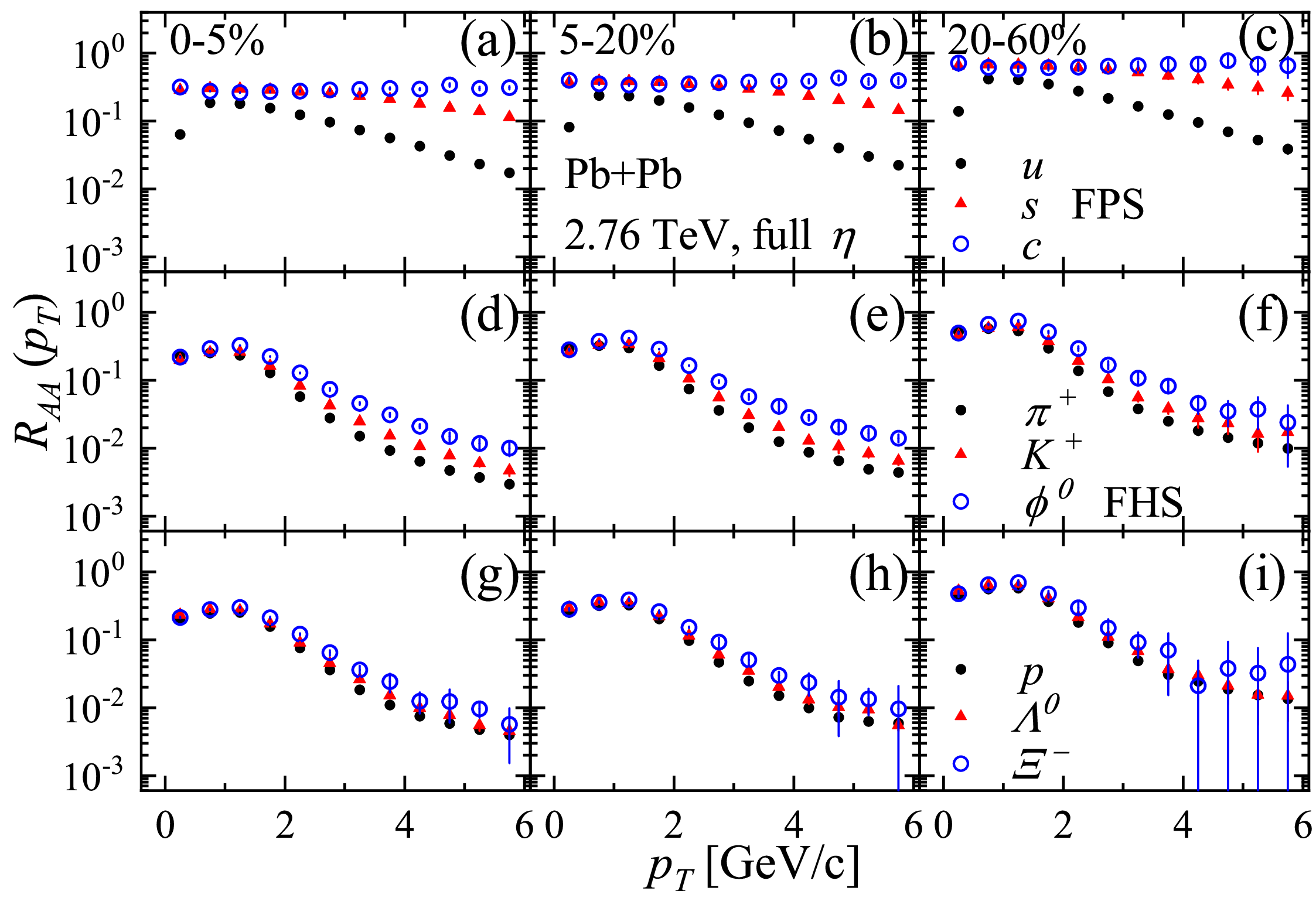}
        \caption{The 
        simulated $R_{AA}$ of quarks (in FPS), mesons 
        (in FHS), and baryons (in FHS) in the different centrality classes of 
        Pb+Pb collisions at $\sqrt{S_{NN}}$=2.76 TeV.}
        \label{fig:raa_diff_cent}
    \end{figure}

    Meanwhile, in Fig.~\ref{fig:raa_diff_cent} we give the simulated $R_{AA}$ 
    of quarks (in FPS), mesons (in FHS) and baryons (in FHS) in 0-5\%, 5-20\%, 
    and 20-60\% centrality classes Pb+Pb collisions at $\sqrt{S_{NN}}$=2.76 
    TeV. The good mass ordering in the region of $p_T>$2 GeV/c is 
    generally held and almost insensitive to the event centrality.

\subsection{Energy dependence} \label{subsec:ener}
\begin{figure}[htbp]
        \centering
        \includegraphics[width=0.5\textwidth]{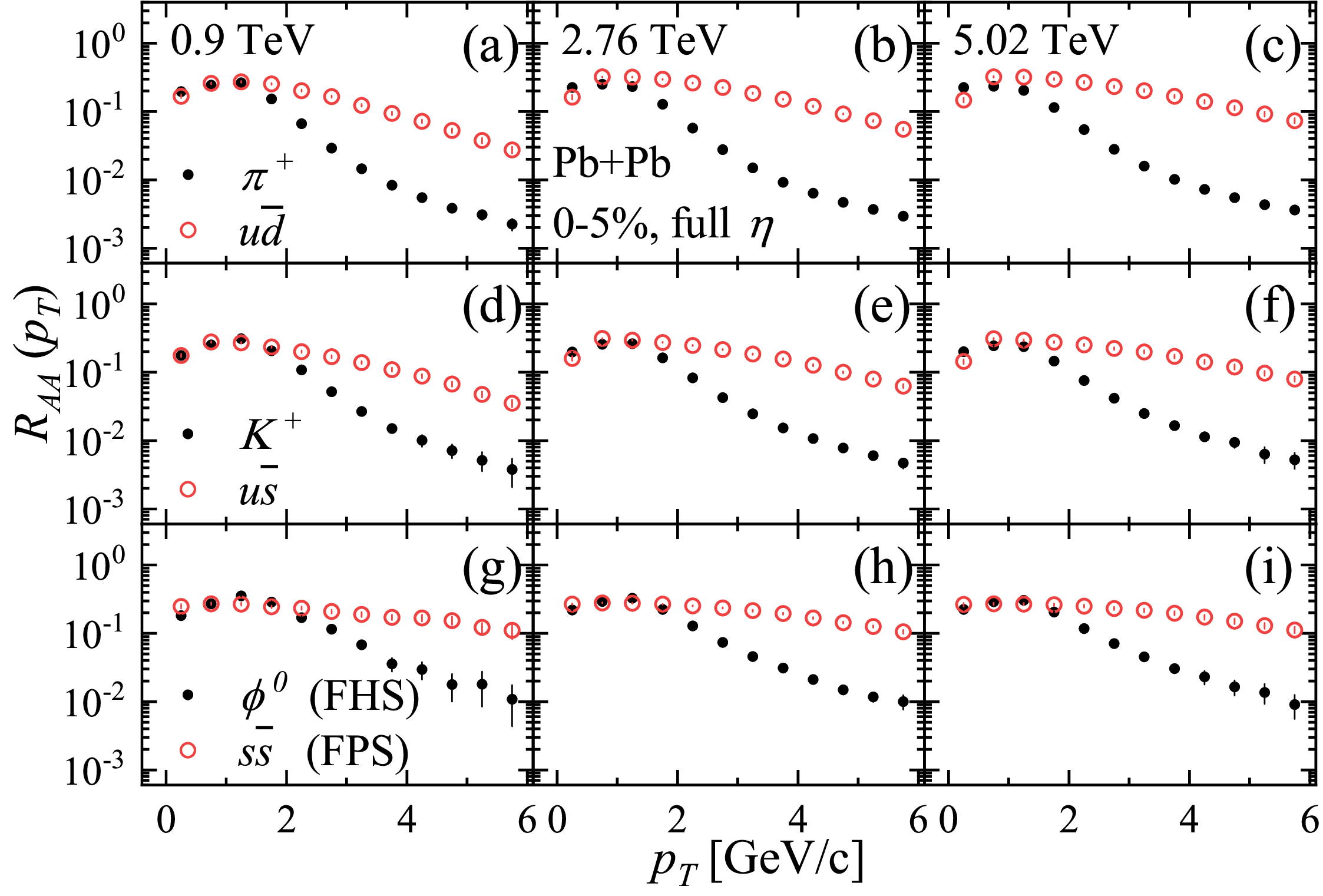}
        \caption{The 
        simulated correspondence between $R_{AA}$ of mesons (in FHS, black 
        solid circles) and their quark component (in FPS, red open circles) in 
        the 0-5\% most central Pb+Pb collisions at different reaction 
        energies.}
        \label{fig:wRaa_mes_diff_ener}
    \end{figure}

\begin{figure}[htbp]
        \centering
        \includegraphics[width=0.5\textwidth]{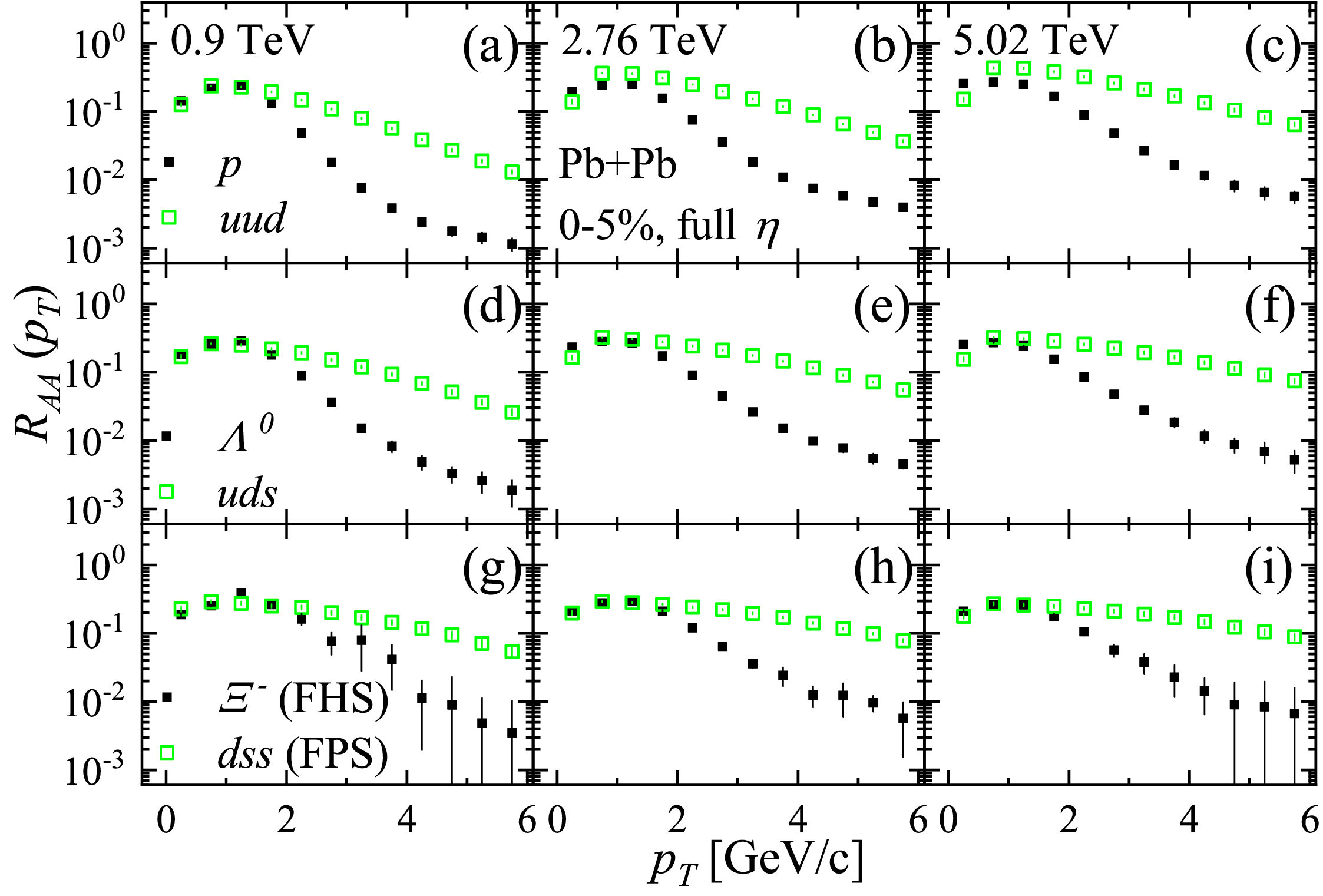}
        \caption{The 
        simulated correspondence between $R_{AA}$ of baryons (in FHS, black 
        solid squares) and their quark component (in FPS, green open squares) in 
        the 0-5\% most central Pb+Pb collisions at different reaction energies.}
        \label{fig:wRaa_bar_diff_ener}
    \end{figure}

    We now study the energy dependence of the 
    correspondence between hadron $R_{AA}$ and its quark component one, 
    and the mass ordering at both the parton and hadron levels. 
    In Figs.~\ref{fig:wRaa_mes_diff_ener} and \ref{fig:wRaa_bar_diff_ener}, 
    we give the $R_{AA}$ of both the hadrons (in FHS) and their quark component 
    (in FPS) in 0-5\% most central Pb+Pb collisions 
    at $\sqrt{S_{NN}}$=0.9, 2.76 and 5.02 
    TeV. The correspondence is well kept at all three reaction energies. 
    However, the discrepancy of $R_{AA}$ between hadron and its quark 
    component is not varying strongly with the collision energies.

\begin{figure}[htbp]
        \centering
        \includegraphics[width=0.5\textwidth]{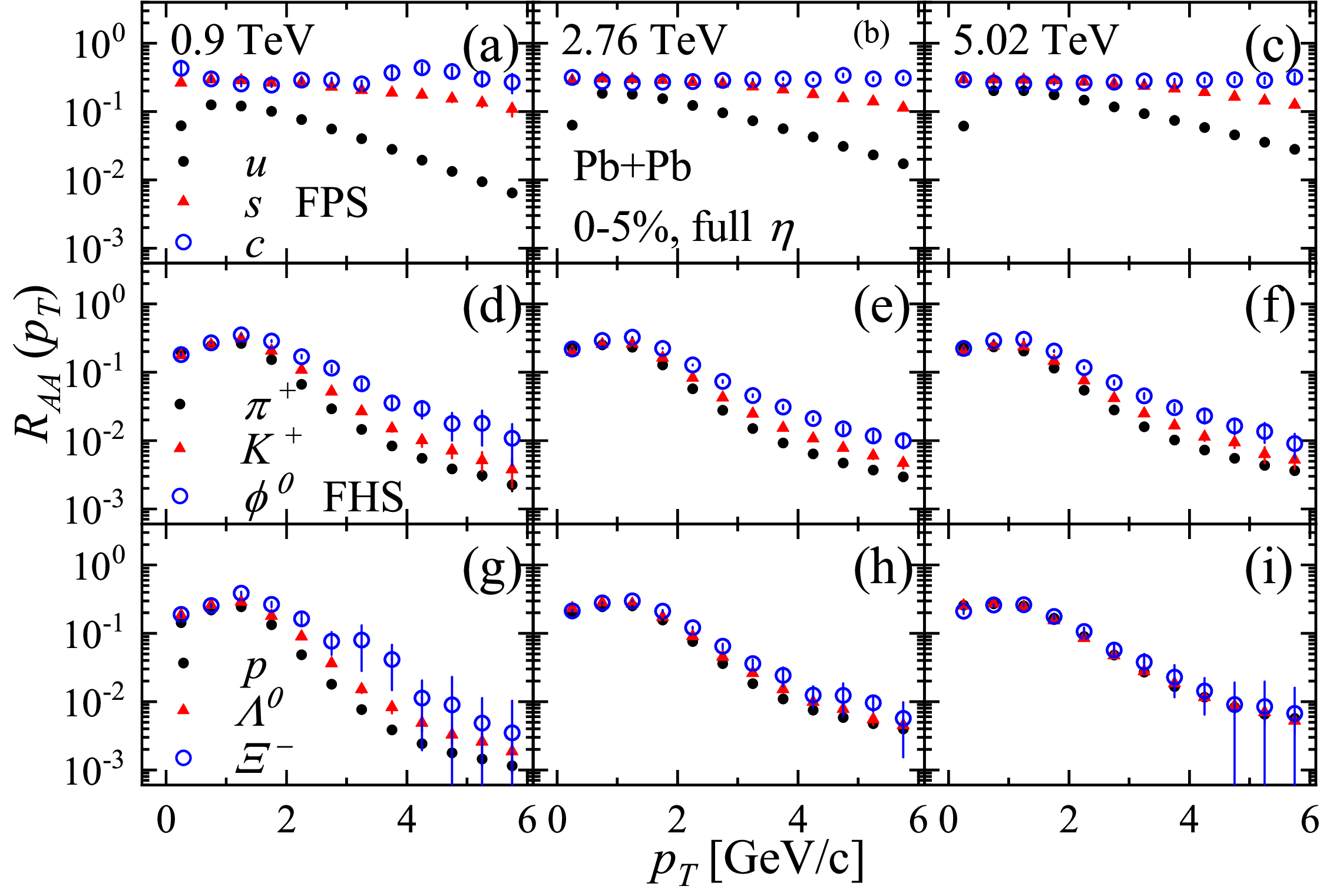}
        \caption{The 
        simulated $R_{AA}$ of quarks (in FPS), mesons 
        (in FHS), and baryons (in FHS) in the 0-5\% most central Pb+Pb 
        collisions at different reaction energies.}
        \label{fig:raa_diff_ener}
    \end{figure}

    In Fig.~\ref{fig:raa_diff_ener} the mass ordering at both the 
    parton and hadron levels are given. Here we can see the mass ordering at 
    hadron level appears in all three reaction energies, like the one at parton 
    level. However, it also seems that the mass ordering at both 
    parton and hadron levels is more pronounced at the lower energy 
    than the higher one. It should be studied further.

\subsection{System size dependence} \label{subsec:syst}
\begin{figure}[htbp]
        \centering
        \includegraphics[width=0.5\textwidth]{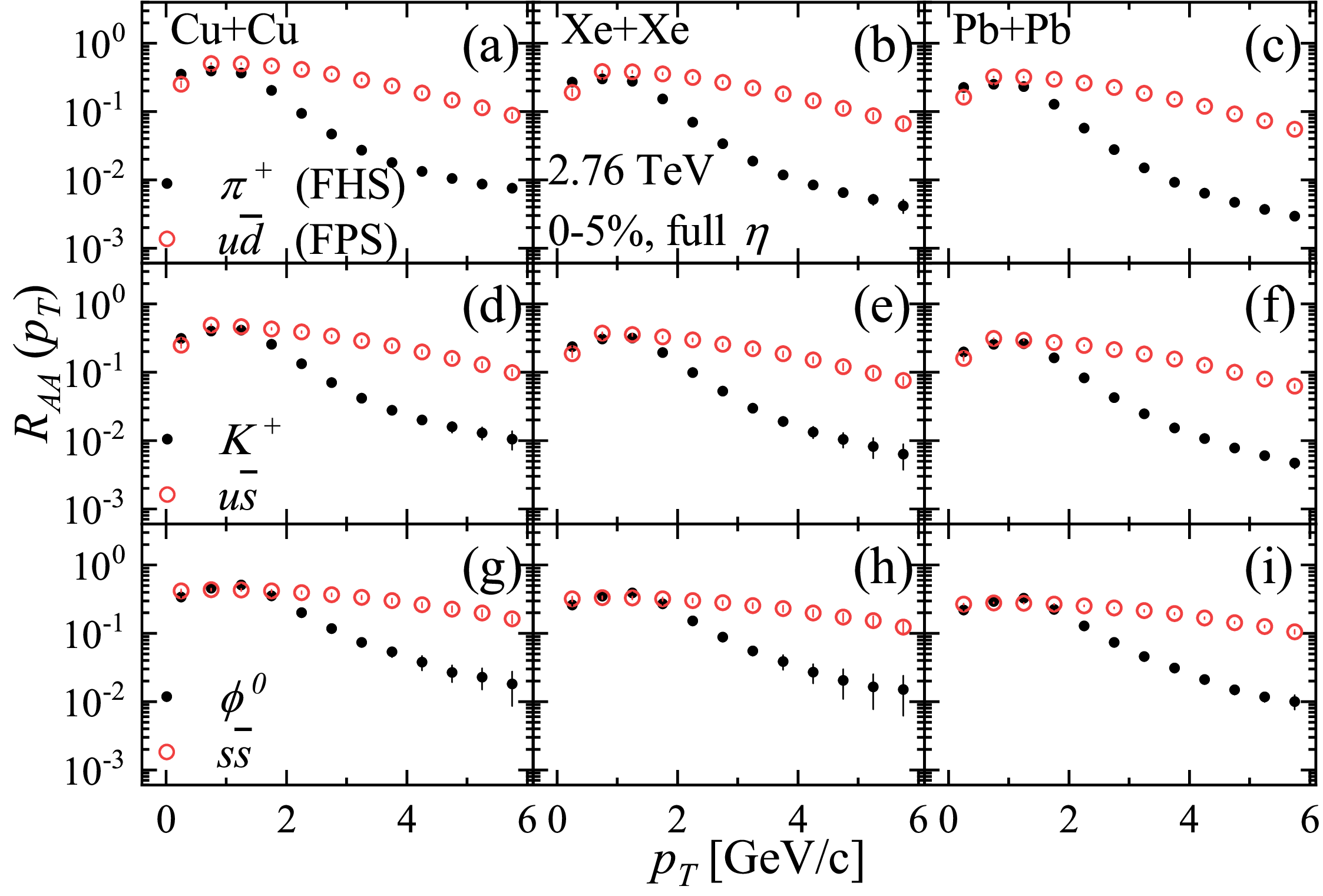}
        \caption{The 
        simulated correspondence between $R_{AA}$ of mesons (in FHS, black 
        solid circles) and their quark component (in FPS, red open circles) in 
        the 0-5\% most central Cu+Cu, Xe+Xe and Pb+Pb collisions at 
        $\sqrt{S_{NN}}$=2.76 TeV.}
        \label{fig:wRaa_mes_diff_syst}
    \end{figure}

\begin{figure}[htbp]
        \centering
        \includegraphics[width=0.5\textwidth]{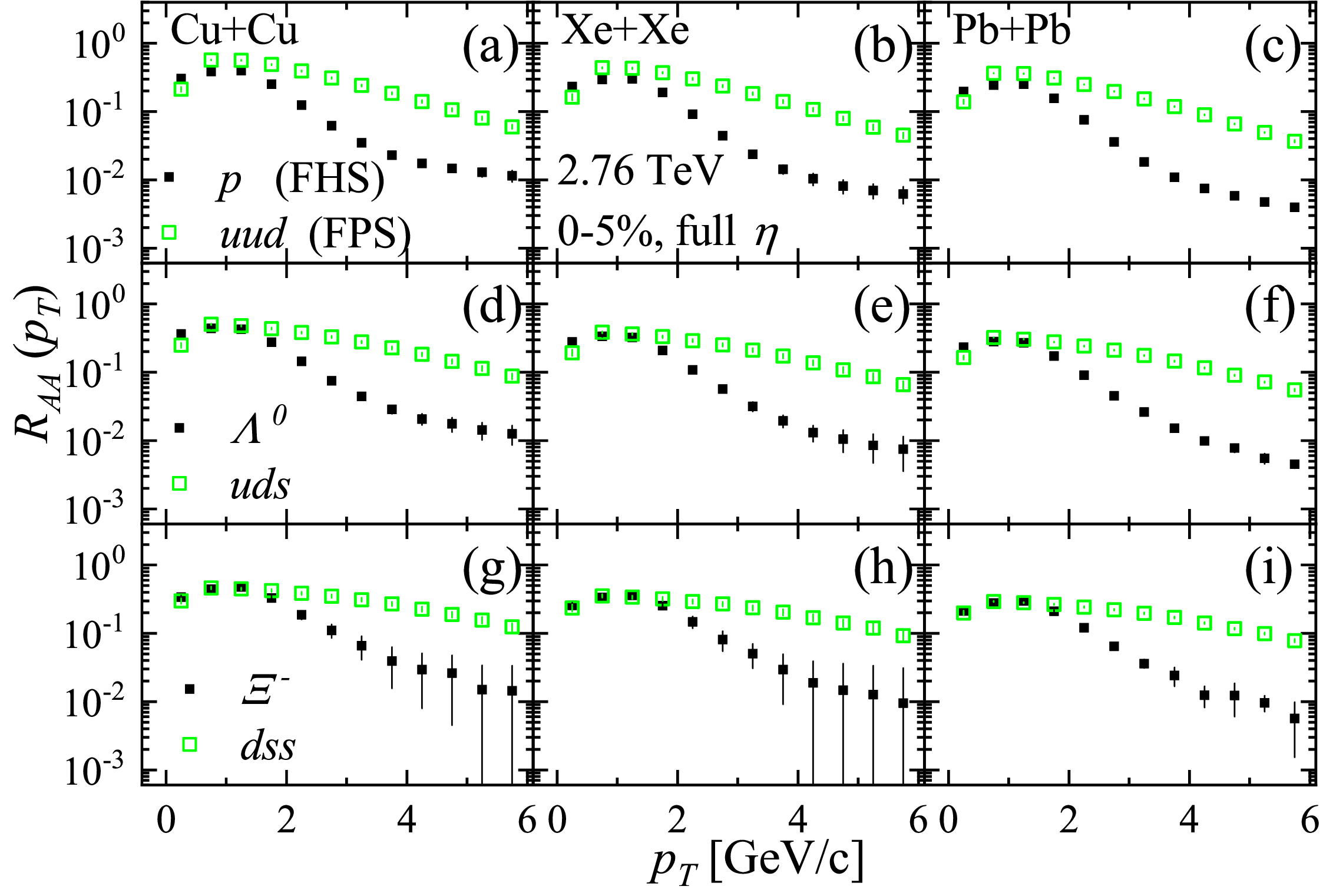}
        \caption{The 
        simulated correspondence between $R_{AA}$ of baryons (in FHS, 
        black solid squares) and their quark component (in FPS, green open 
        squares) in the 0-5\% most central Cu+Cu, Xe+Xe and Pb+Pb collisions 
        at $\sqrt{S_{NN}}$=2.76 TeV.}
        \label{fig:wRaa_bar_diff_syst}
    \end{figure}

\begin{figure}[htbp]
        \centering
        \includegraphics[width=0.5\textwidth]{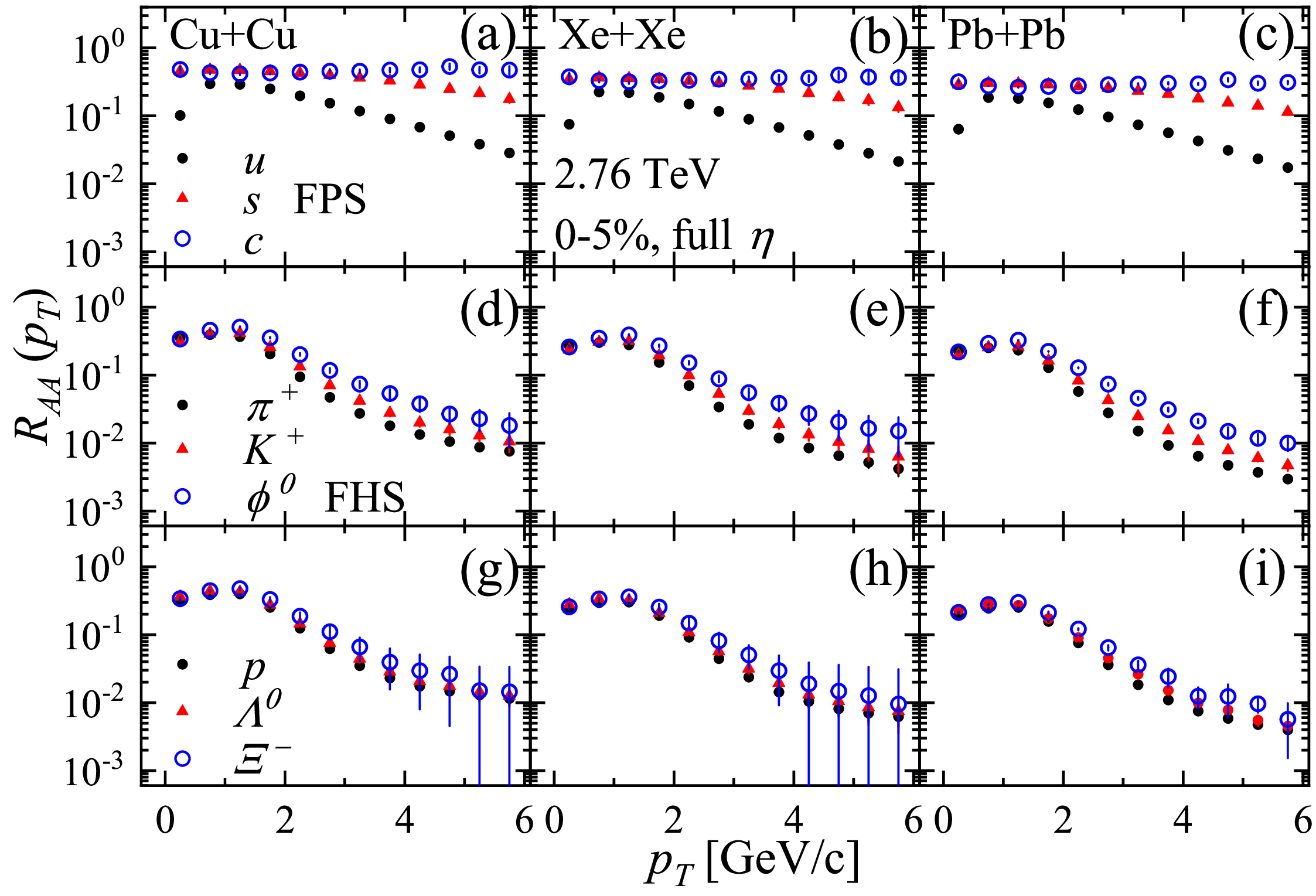}
        \caption{The 
        simulated $R_{AA}$ of quarks (in FPS), mesons 
        (in FHS), and baryons (in FHS) in the 0-5\% most central 
        Cu+Cu, Xe+Xe and Pb+Pb collisions at $\sqrt{S_{NN}}$=2.76 TeV.}
        \label{fig:raa_diff_syst}
    \end{figure}

    In Figs.~\ref{fig:wRaa_mes_diff_syst} (meson) and 
    \ref{fig:wRaa_bar_diff_syst} (baryon), we give the simulated hadron 
    $R_{AA}$ and its quark component $R_{AA}$ in 0-5\% most central Cu+Cu, 
    Xe+Xe and Pb+Pb collisions at $\sqrt{S_{NN}}$=2.76 TeV. We present the 
    mass ordering at both the parton and hadron levels in 
    Fig.~\ref{fig:raa_diff_syst} for the same reaction systems above. These 
    figures show again that, the correspondence between hadron and its quark 
    component in $R_{AA}$ as well as the $R_{AA}$ mass ordering are well kept.

    As the amount of interacting matter increases with the system size, 
    the particle propagating length is longer in larger collision system. 
    Hence the larger the system size, the more energy losses there
    are~\cite{Wang:2004dn}. Consequently, the suppression of $R_{AA}$ would 
    decrease with the collision system size, as shown in 
    Figs.~\ref{fig:wRaa_mes_diff_syst},~\ref{fig:wRaa_bar_diff_syst} 
    and~\ref{fig:raa_diff_syst} from left to right. 

\section{Summary} \label{sec:sum}
    In summary, via the parton and hadron cascade model PACIAE, we study the 
    connection (correspondence) between hadron nuclear modification factor and 
    its quark component one, as well as the flavor (mass) ordering at both 
    parton and hadron levels. Meanwhile, how the above two physical phenomena 
    change with the (pseudo-)rapidity, centrality, reaction energy and the 
    collision system size are investigated systematically.

    Generally speaking, the correspondence between hadron and its quark 
    component and the mass ordering at both the parton and hadron level in 
    nuclear modification factors are held, irrespective of the rapidity, 
    centrality, reaction energy, and the collision system size.

    For the correspondence, the $R_{AA}$ of the hadron is always less than that
    of its quark component in the $p_T$ region above 2 GeV/c. The discrepancy 
    between them becomes more pronounced in a wider $\eta$ range. However, 
    this behavior do not noticeably show in the different centrality, reaction 
    energy and the system size studies. The mass ordering is easier to 
    distinguish in the wider $\eta$ and the lower reaction energy, while it 
    does not exhibit the clear dependences on the centrality and the collision 
    system size.

    We note that, the figures displaying the correspondence between hadron and 
    its quark component in $R_{AA}$, show a similar phenomenon: The 
    discrepancy between hadron $R_{AA}$ and its quark component $R_{AA}$ in 
    low $p_T$ region is less than the one in middle and/or high $p_T$ region. 
    It is not understood very well yet and has to leave it to the next study.

    The nuclear modification factors at hadron level show very well mass 
    ordering, like the one at parton level. Its clear observation is relevant 
    to the relative mass discrepancy among the selected candidates. Larger 
    relative mass discrepancy among the candidates leads to more significant 
    mass ordering.

    Of course, the correspondence between hadron and its quark component and 
    the hadronic mass ordering in the nuclear modification factor should be 
    studied further, both theoretically and experimentally. In the next work, 
    we would consider the open-charm and/or the open-bottom 
    heavy-hadron~\cite{Andronic:2015wma} as the candidates.

\begin{acknowledgments}
    We thank Dr. Chun-Bin Yang for discussions. 
    This work was supported by the National Natural 
    Science Foundation of China (11775094, 11905188, 11775313, 11905163), the 
    Continuous Basic Scientific Research Project (No.WDJC-2019-16), National 
    Key Research and Development Project (2018YFE0104800) and by the 111 
    project of the foreign expert bureau of China.
\end{acknowledgments}

%

\end{document}